\newif\ifNotes\Notestrue
\newif\ifAnon\Anonfalse
\newif\ifDraft\Draftfalse
\newif\ifTor\Tortrue
\newif\ifNDSS\NDSSfalse
\newif\ifCCS\CCSfalse
\newif\ifUSENIX\USENIXtrue
\newif\ifOld\Oldtrue
\newif\ifArxiv\Arxivfalse
\newif\ifCamera\Camerafalse

\Arxivtrue

\ifCamera
  \Notesfalse
  \Anonfalse
  \Draftfalse
  \Arxivfalse
\fi

\ifCCS
  \ifAnon
    \documentclass[sigconf, anonymous]{acmart}
  \else
    \documentclass[sigconf]{acmart}
  \fi
  \usepackage{appendix}
\fi
\ifNDSS
  \documentclass[conference]{IEEEtran}
  \usepackage{color}
  \usepackage[numbers,sort]{natbib}
\fi
\ifUSENIX
  \documentclass[letterpaper,twocolumn,10pt]{article}
  \usepackage{usenix2019,epsfig,endnotes}
  \ifCamera
  \else
  \fi
  \usepackage{color}
  \usepackage[numbers,sort]{natbib}
  \usepackage{appendix}
\fi

\usepackage{booktabs} 
\usepackage{xspace}
\usepackage[hyphens]{url}
\usepackage{hyperref}
\usepackage[flushleft]{threeparttable}
\usepackage{pifont}
\usepackage{courier}
\usepackage{graphicx}
\usepackage{array}
\usepackage{soul}
\usepackage[table]{xcolor}

\usepackage{enumitem}

\ifDraft
  \usepackage{draftwatermark}
  \definecolor{watermarkcolor}{rgb}{0.8,0.8,1}
  \SetWatermarkText{DRAFT}
  \SetWatermarkColor{watermarkcolor}
\fi

\definecolor{linkcolor}{rgb}{0.65,0,0}
\definecolor{citecolor}{rgb}{0,0.4,0}
\definecolor{urlcolor}{rgb}{0,0,0.65}
\hypersetup{hyperindex=true,pdfpagemode=UseNone,pdfstartview=FitH,colorlinks=true, linkcolor=linkcolor, urlcolor=urlcolor, citecolor=citecolor}

\providecommand{\tabularnewline}{\\}

\usepackage{color}

\usepackage[english]{babel}
\usepackage{blindtext}
\usepackage{tikz}

\newcommand{\swallow}[1]{}
\ifNotes
  \newcommand{\colorcomment}[2]{\leavevmode\unskip\space{\color{#1}[#2]}\xspace}
\else
  \newcommand{\colorcomment}[2]{\leavevmode\unskip\relax}
\fi

\ifArxiv
\newcommand{\noArxiv}{\errmessage{Remove comments for arXiv}}
\else
\newcommand{\noArxiv}{\relax}
\fi

\newcommand{\tpm}{$\pm$}
\newcommand{\btpm}{\boldmath$\pm$}
\newcommand{\tha}[2]{\multicolumn{1}{#1}{#2}}

\ifOld
  \newcommand{\old}[1]{\noArxiv\leavevmode\unskip\relax}
  
\else
  \newcommand{\old}[1]{\noArxiv\leavevmode\unskip\relax}
  
\fi

\newcommand{\parhead}[1]{\vspace{3pt plus 1pt minus 1pt}\par\noindent\textbf{#1}\hspace{.75em plus .5em minus .5em}}

\newcommand{\pp}{Prime+\allowbreak Probe\xspace}

\usepackage{cleveref}
\crefformat{section}{#2Section~#1#3}
\Crefformat{section}{#2Section~#1#3}
\crefname{figure}{Figure}{Figures}
\Crefname{figure}{Figure}{Figures}
\crefformat{figure}{#2Figure~#1#3}
\Crefformat{figure}{#2Figure~#1#3}
\crefname{table}{Table}{Tables}
\Crefname{table}{Table}{Tables}
\crefformat{table}{#2Table~#1#3}
\Crefformat{table}{#2Table~#1#3}
\crefname{appsec}{Appendix}{Appendices}
\Crefname{appsec}{Appendix}{Appendices}
\crefformat{appsec}{#2Appendix~#1#3}
\Crefformat{appsec}{#2Appendix~#1#3}

\ifCCS
  \newcommand{\authorlist}[1]{#1}
  \newcommand{\nextauthor}{\relax}
  \newcommand{\myAuthor}[3]{
    \author{#1}
    \affiliation{#2}
    \email{#3}
  }
\fi
\ifNDSS
  \ifAnon
    \newcommand{\authorlist}[1]{\relax}
  \else
    \newcommand{\authorlist}[1]{\author{#1}}
  \fi
  \newcommand{\nextauthor}{\and}
  \newcommand{\myAuthor}[3]{
    \IEEEauthorblockN{#1}
    \IEEEauthorblockA{#2 \\ 
    #3}
  }
\fi 
\ifUSENIX
  \ifAnon
    \newcommand{\authorlist}[1]{\relax}
  \else
    \newcommand{\authorlist}[1]{\author{#1}}
  \fi
  \newcommand{\nextauthor}{\and}
  \newcommand{\myAuthor}[3]{
    {\rm #1} \\
    #2\\
    #3
  }
\fi

\ifCCS
  \renewcommand\footnotetextcopyrightpermission[1]{} 
  \acmDOI{}

  \acmISBN{}

  \acmConference[Submission \#756]{ACM}{May 2018}{USA}  
  \acmYear{1997}

\fi

\begin{document}

\title{Robust Website Fingerprinting Through the Cache Occupancy Channel}

\authorlist{
  \myAuthor{Anatoly Shusterman}{Ben-Gurion University  of the Negev}{shustera@post.bgu.ac.il}\\
  \nextauthor
  \myAuthor{Lachlan Kang}{University of Adelaide}{lachlan.kang@adelaide.edu.au}\\
  \nextauthor
  \myAuthor{Yarden Haskal}{Ben-Gurion Univ. of the Negev}{yardenha@post.bgu.ac.il}\\
  \nextauthor
  \myAuthor{Yosef Meltser}{Ben-Gurion Univ.  of the Negev}{yosefmel@post.bgu.ac.il}
  \nextauthor
  \myAuthor{Prateek Mittal}{Princeton University}{pmittal@princeton.edu}\\
  \nextauthor
  \myAuthor{Yossi Oren}{Ben-Gurion Univ.  of the Negev}{yos@bgu.ac.il}
  \nextauthor
  \myAuthor{Yuval Yarom}{University of Adelaide and Data61}{yval@cs.adelaide.edu.au}
}

\ifUSENIX
\maketitle
\fi

\begin{abstract} 
  Website fingerprinting attacks, which use statistical analysis on network
  traffic to compromise user privacy, have been shown to be effective even if
  the traffic is sent over anonymity-preserving networks such as Tor.  The
  classical attack model used to evaluate website fingerprinting attacks
  assumes an \emph{on-path adversary}, who can observe all traffic traveling
  between the user's computer and the secure network. 
  
  In this work we investigate these attacks under a different attack model, in
  which the adversary is capable of sending a small amount of malicious JavaScript code
  to the target user's computer.
  The malicious code mounts a
  cache side-channel attack, which exploits the effects of contention on the
  CPU's cache, to identify \emph{other} websites being browsed.  
  The effectiveness of this
  attack scenario has never been systematically analyzed, especially in the
  open-world model which assumes that the user is visiting a mix of both
  sensitive and non-sensitive sites.
  
  We show that cache website fingerprinting attacks in JavaScript
  are highly feasible.
  Specifically, we use machine learning
  techniques to classify traces of cache activity.  Unlike prior works, which
  try to identify cache conflicts, our work measures the overall occupancy of
  the last-level cache.  We show that our approach achieves high classification
  accuracy in both the open-world and the closed-world models. We further show
  that our attack is more resistant than network-based fingerprinting to the
  effects of response caching, and that our techniques are resilient both to
  network-based defenses and to side-channel countermeasures introduced to
  modern browsers as a response to the Spectre attack. 
  To protect against
  cache-based website fingerprinting, new defense mechanisms must be introduced
  to privacy-sensitive browsers and websites.
  We investigate one such mechanism, and show that generating artificial cache
  activity reduces the effectiveness of the attack and completely eliminates it
  when used in the Tor Browser.

  \end{abstract}

%

%
\ifCCS

  \keywords{Website fingerprinting, cache attacks, Tor, deep learning}

  \maketitle
\fi

\section{Introduction}\label{s:introduction}
Over the last decades the World Wide Web  has grown from an
academic exercise to a communication tool that encompasses all aspects
of modern life.
Users use the web to acquire information, manage their finances, 
conduct their social life, and more.
This shift to the so called virtual life has resulted in new challenges to users' privacy.
Monitoring the online behavior of users may reveal personal or sensitive
information about the users,
including information such as sexual orientation or political beliefs and affiliations.

Several tools have been developed to protect the online privacy of users
and hide information about the websites they visit~\cite{ReiterR98,DingledineMS04,Dai98}.
Prime amongst these is the Tor network~\cite{DingledineMS04}, 
an overlay network of collaborating servers, called \emph{relays}, that anonymously
forward Internet traffic between users and web servers.
Tor encrypts the network traffic of all of the users,
and transmits it between relays in a way that prevents external observers
from identifying the traffic of specific users.
In addition to the network itself, the Tor Project also provides the 
\emph{Tor Browser}~\cite{TorBrowser}, a
modified version of the Mozilla Firefox web browser,
that further protects users by disabling features that may allow web
sites to track the users.

Past research has demonstrated that encrypting traffic is not sufficient
for protecting the privacy of the users~\cite{RimmerPJVJ18,GongBKS12,
PanchenkoLPEZHW16,WangG16,HayesD16,JuarezAADG14,YanK18,JansenJGED18,LiGH18,
LuCC10,WangG13,CaiZJJ12,HerrmannWF09,PanchenkoNZE11,Hintz02}.
Observable patterns in the metadata of encrypted traffic, 
specifically, the size of the transmitted data, its direction, and its timing,
may reveal the web page that the user is visiting.
Applying such \emph{website fingerprinting} techniques to
Tor traffic results in a
success rate of over 90\% in identifying the websites
that a user visits over Tor~\cite{RimmerPJVJ18}.\footnote{\emph{Website fingerprinting}
is a misnomer. Fingerprinting identifies individual web pages rather than sites.
Following this misnomer, in this work we use the term \emph{website} to refer to
specific pages, typically the homepage of the site.}

In this paper, we focus on an alternative attack model of exploiting micro-architectural side-channels, a less explored option for website fingerprinting. 
The attack model assumes a victim that visits a
web site under the attacker's control.
The web site monitors the state of the victim computer's cache,
and uses that information to infer the victim's web activity in
other tabs of the same browser, or even in other browsers.

Because the attack observes the internal state of the target PC, rather than the network
traffic.
It offers the potential of overcoming traffic shaping, often proposed
as a defense for website fingerprinting~\cite{CaiNWJG14, NithyanandCJ14, CaiNJ14, WangG17, CherubinHJ17}.
Similarly, the attack may be applicable in scenarios where network-based fingerprinting
is known to be less effective, such as when the browser caches the contents
of the website~\cite{HerrmannWF09}.

We note that the malicious web site does not need to be fully under the control
of the attacker.
The attacker only needs to be able to inject JavaScript code via the
web site to the victim's browser.
This can be done, for example, through a
malicious advertisement or pop-up window.  Alternatively, documents released by former NSA
contractor Edward Snowden indicate that some nation-state agencies have the
operational capability to exploit this vector on a wide scale.  In March 2013 the
German magazine Der Spiegel reported on the existence of a tool called
\textsc{quantuminsert}, which the GCHQ and the NSA could use to inject malicious code
to any website~\cite{Spiegel13}.  The Der Spiegel claims that the GCHQ 
successfully used this tool to attack the computers of employees at the
partly-government-held Belgian telecommunications company Belgacom, and that the NSA used
the same technology to target high-ranking members of the Organization of the
Petroleum Exporting Countries (OPEC) at the organization's Vienna headquarters.
Finally, malicious advertisements are a viable
option for injecting cache side-channel attacks to browsers~\cite{GenkinPTY18}.

For a small number of websites, under the closed-world model,
\citet{OrenKSK15} show the possibility of fingerprinting via malicious JavaScript code.
However, beyond showing the ability to distinguish between a handful of websites,
their work does not provide an analysis of the effectiveness of
the technique.
Furthermore, following the disclosure of the Spectre and the Meltdown attacks, which
can also be potentially delivered via malicious JavaScript
injection~\cite{lipp2018meltdown,kocher2018spectre}, major vendors 
deployed defenses against browser-borne side-channel attacks.  In particular, all
modern browsers have reduced the resolution of the JavaScript time function,
\texttt{performance.now()}, by several orders of magnitude~\cite{Wagner18,Pizlo18},
making it difficult to tell apart cache hits and cache misses.  
Traditionally, cache attacks require high-resolution timers, and while
mechanisms to generate such timers in web browsers have been 
published~\cite{SchwarzMGM17,GrasRBBG17,KohlbrennerS16},
it is not clear that these can be used for website fingerprinting.

Thus, in this paper we ask:
\emph{Are cache-based attacks a viable option for website fingerprinting?}

\subsection*{Our Contribution}
We answer this question in the affirmative.
  We design and implement a cache-based website fingerprinting attack,
  and evaluate it in both the closed-world and the open-world models.
  We show that in both models 
  our JavaScript-based attacker achieves high fingerprinting accuracy
  even when executed on modern mainstream browsers that include all
  recently introduced countermeasures for side-channel (Spectre) attacks.
  Even when taking these countermeasures to the extreme,
  as is done in the Tor Browser, our attack remains effective,
  although with a drop in accuracy.

  Our attack consists of collecting traces of cache \emph{occupancy}
  while the browser downloads and renders web sites.
  Adapting the techniques of \citet{RimmerPJVJ18}, 
  we use deep neural networks to analyze and to classify the collected
  traces.
  By focusing on cache occupancy rather than on activity 
  within specific cache sets, our attack avoids the need for 
  high resolution timers required by prior cache-based attacks.
  Furthermore, because our technique does not depend on the
  layout of the cache, it can overcome proposed countermeasures
  that randomize the cache layout~\cite{Qureshi18,LiuL14,WangL07}.
   
  We investigate the source of the information in the 
  cache occupancy traces and show that they contain
  information from both the networking activity and the rendering
  activity of the browser.
  Using information from the rendering activity allows our attack
  to remain effective even in scenarios that thwart network-based fingerprinting,
  such as when the browser retrieves data from its response cache and
  not from the network, or when the network traffic is shaped.

  Finally, we investigate a potential countermeasure that introduces
  a high level of activity into the last level cache.  We show
  that the countermeasure reduces the success rate
  of the attack.  In particular, the noise completely masks the activity
  of the Tor Browser, reducing the attack accuracy to that of a 
  random guess.  
  This countermeasure results in a mean slowdown of 5\% for
  CPU benchmarks, which we consider reasonable when visiting
  privacy-sensitive web sites.

More specifically, we make the following contributions:
\begin{itemize}[itemsep=1ex,topsep=1ex,parsep=0pt,partopsep=0pt]
  \item 
    We design and implement the cache occupancy side-channel attack,
    a cache-based side channel attack technique 
    which can operate with the low timer resolution supported in 
    modern JavaScript engines.
    Our attacks only require a sampling rate
    six orders of magnitude \emph{lower} than required for the prior attacks
    of Oren et al.~\cite{OrenKSK15}
    (\cref{s:memorygrammer}).
  \item We evaluate the use of two machine learning techniques, CNN and LSTM,
    for fingerprinting websites 
    based on the cache activity traces collected while loaded by the browsers (\cref{s:ml}).
  \item We show that cache-based fingerprinting has high accuracy in both the closed-
    and the open-world models, under a variety of operating systems and browsers
    (\cref{s:results}).
  \item We evaluate both fingerprinting methods without deleting the
    browser response cache, and show that while the accuracy of network-based
    fingerprinting drops significantly, the accuracy of cache-based fingerprinting
    is not affected  (\cref{s:keep-cache}). 
  \item We show that cache-based fingerprints contain information both from the
    network activity and from the rendering activity of the target device. Therefore,
    cache-based fingerprinting maintains a high accuracy even in the presence of traffic
    molding countermeasures which force a constant bit rate on network traffic
    (\cref{s:worst-case-molding}).  
  \item We design and evaluate a countermeasure that introduces noise in the cache.
    The countermeasure is applicable from both native code
    and from JavaScript, completely blocks the attack on the Tor Browser,
    and only causes a small performance degradation on CPU-bound workloads
    (\cref{s:countermeasures}).
\end{itemize}

\section{Background}
\subsection{Tor}
Tor~\cite{DingledineMS04}, is a collection of collaborating servers called \emph{relays},
designed to provide privacy for network communication.
Tor aims to protect users from \emph{on-path} adversaries that can observe the network
traffic.
In this scenario, a user uses a PC to browse the web, 
and an adversary positioned between the user's PC and the destination web server
captures the information that the user exchanges with the web server.

A common protection for such an attack model is to use encryption, e.g., using
protocols such as TLS~\cite{RFC5246} which underlies the security of the HTTPS
scheme~\cite{RFC2818}.  However, this solution only protects the contents of the
communication, leaving the identity of the communicating parties exposed to the
adversary. Knowing that users merely connected to a certain sensitive
website may be enough to incriminate them, even if the actual data exchanged over the
secure connection is not known. This risk became a reality in 2016, as tens of
thousands of individuals were persecuted by the Turkish government for accessing the
domain \texttt{bylock.net}~\cite{Koskal18}.

The main aim of Tor is thus to protect the identity of the communicating parties.
Tor achieves this protection by forwarding the users' communication through
a \emph{circuit} consisting of a few (typically three) Tor relays.
The user encrypts the network traffic with multiple layers of encryption,
and each relay in the circuit decrypts a successive layer to find out where
to forward the traffic.
See~\citet{DingledineMS04} for further information.

 \subsection{Website Fingerprinting Attacks and Defences}
In the conventional attack model of a network-level attacker, much previous work has
demonstrated the ability of an adversary to make probabilistic inferences about
users' communications via statistical analysis, even if these communications are in
their encrypted form. These works have investigated both the selection of features
(such as packet sizes, packet timings, direction of communication), as well as the
design of classifiers (such as support vector machines, random forests, Naive Bayes)
to make accurate predictions~\cite{RimmerPJVJ18,GongBKS12,
PanchenkoLPEZHW16,WangG16,HayesD16,JuarezAADG14,JansenJGED18,YanK18,LiGH18,
LuCC10,WangG13,CaiZJJ12,HerrmannWF09,PanchenkoNZE11,Hintz02}.  In response, several
defense mechanisms have been proposed in the
literature~\cite{CaiNWJG14,NithyanandCJ14,CaiNJ14,WangG17,CherubinHJ17}.  The common
idea behind these defenses is to inject random delays and spurious cover traffic to
perturb the traffic features and therefore obfuscate users' communications. A common
point of all of these defenses is a typical trade-off between latency/bandwidth and
privacy, and thus they face deployment hurdles. \citet{RimmerPJVJ18} have recently
proposed a family of classifiers based on deep learning algorithms such as SDAE, CNN
and LSTM, which operate on the raw network traces and are therefore less sensitive to
ad-hoc defenses against particular traffic features.

\subsection{Cache Side-Channel Attacks}\label{s:cacheSC}
When programs execute on a processor, they share the use of micro-architectural
components such as the cache.
This sharing may result in unintended communication
channels, often called \emph{side channels}, between programs~\cite{Hu92,GeYCH18},
which may be used to leak secret information.
In particular, cache-based attacks, which exploit
contention on one of the processor's caches, can leak secrets
such as cryptographic keys~\cite{OsvikST06,Percival05,GarciaBY16,TsunooSSSM03,AciicmezBG10}, 
keystrokes~\cite{GrussSM15},
address layout~\cite{GrussMFLM16,EvtyushkinPA16,GrasRBBG17}, etc.

\parhead{Cache Operation.}
Caches bridge the speed gap between the faster processor and the slower memory.
The cache is a small bank of memory, which stores the contents of recently
accessed memory locations.
Most caches in modern processors are \emph{set associative}.
The cache is divided into partitions called \emph{sets}.
Each memory location maps to a single set and can only be cached in the set it maps to.
When the processor needs to access a specific memory location,
it successively searches in a hierarchy of caches.
In a \emph{cache hit}, when the contents of the required address is found in the cache,
access is performed on the cached contents.
Otherwise, in a \emph{cache miss}, the process repeats on the next cache level.
A miss on the last-level cache (LLC) results in a time-consuming access to the RAM.

\parhead{The \pp Technique.}
Past cache-based attacks from web browsers~\cite{OrenKSK15,GenkinPTY18} employ
the \emph{\pp} technique~\cite{OsvikST06,Percival05},
which exploits  the set-associative structure.
Each round of attack consists of three steps.
In the first step, the cache is \emph{primed},
i.e., the attacker completely fills some of the cache sets with its own data.
The attacker then waits some time to allow the victim to execute.
Finally, the attacker \emph{probes} the cache by measuring 
the time it takes to access the previously-cached data in each of the sets.
If the victim accesses memory locations that map to a monitored cache set,
the victim's memory contents will replace the attacker contents in the cache.
Hence, the attacker will need to retrieve the data from lower levels in the
hierarchy, increasing the access time to its data.
\pp has been used for attacks on data~\cite{OsvikST06,Percival05} 
and instruction~\cite{AciicmezBG10,Aciicmez07} caches,
as well as for attacks on the LLC~\cite{LiuYGHL15,IrazoquiES15}.
It has been shown practical in multiple settings, 
including across different virtual machines in cloud environments~\cite{InciGIES16}
and from mobile code~\cite{OrenKSK15,GenkinPTY18}.

\parhead{Countermeasures in JavaScript.}
The time difference between the latencies of a memory access and cache access
is on the order of 0.1\,$\mu$s.  To distinguish
between cache hits and misses, cache attacks typically require a high
resolution timer.  Following the publication of the first demonstration of a cache
attack in JavaScript~\cite{OrenKSK15}, some browsers started reducing the resolution
of the timers they provide as a countermeasure for cache side channel attacks.  This
approach had become wide-spread after the disclosure of the Spectre
attack~\cite{kocher2018spectre}, and now all mainstream browsers incorporate this
countermeasure.  Furthermore, while non-traditional timers in browsers have been
identified~\cite{SchwarzMGM17,KohlbrennerS16,FrigoGBR18}, 
browsers and extensions have since disabled many of the features that allow
sub-microsecond resolution~\cite{Mozilla18,Pizlo18,SchwartzLG18}. An extreme case of this
behavior can be found in the Tor Browser, which restricts the timer resolution to
100\,ms, or 10\,Hz. 

Several of the previously discovered timers rely on browser features that 
are accessible from JavaScript.
These are not accessible in environments such as Cloudflare Workers~\cite{Bloom18},
which rely on the absence of high-resolution timers
to protect against timing attacks~\cite{Varda18}.

\subsection{Related Work}\label{s:relatedWork}
\begin{table*}
\caption{Related work on website fingerprinting based on local side channels.
\label{t:related-work}}

\begin{center}
\resizebox{\textwidth}{!}{%
\begin{centering}
  \begin{tabular}{lp{2.2in}llr}
\toprule 
     &        &              &              & Sampling\tabularnewline
Work & Target & Side Channel & Attack Model & rate [Hz]\tabularnewline
\midrule 
 
Clark et al., 2013 \cite{ClarkMRSFX13} & Chrome (Mac, Win, Linux) & Power consumption & Hardware & $250000$\tabularnewline
 
Yang et al., 2017 \cite{YangGZFB17} & Multiple smartphones & Power consumption & Hardware & $200000$\tabularnewline
 
Lifshits et al., 2018 \cite{LifshitsFHHPTS18} & Android Browser, Chrome Android & Power consumption & Hardware & $1000$\tabularnewline
 
Jana and Shmatikov, 2012 \cite{JanaS12a} & Chrome Linux, Firefox Linux, Android Browser (VM) & App memory footprint & Native code & $100000$\tabularnewline
 
Lee et al., 2014 \cite{LeeKKK14} & Chromium Linux, Firefox Linux & GPU memory leaks & Native code & N/A\tabularnewline
 
Spreitzer et al., 2016 \cite{SpreitzerGKM16} & Chrome Android, Android Browser, Tor Android & Data-Usage Statistics & Native code & 20\textendash 50\tabularnewline
 
G\"ulmezoglu et al., 2017 \cite{GulmezogluZES17} & Chrome Linux (Intel and ARM), Tor Linux & Performance counters & Native code & $10000$\tabularnewline
 
Oren et al, 2015 \cite{OrenKSK15} & Safari MacOS, Tor MacOS & Last-level cache & JavaScript & $10^{8}$\tabularnewline
 
Booth, 2015 \cite{NotSoIncognito}  & Chrome (Mac, Win, Linux), Firefox Linux & CPU activity & JavaScript & $1000$\tabularnewline
 
Kim et al., 2016 \cite{KimL016} & Chromium Linux, Chrome (Win, Android) & Quota Management API & JavaScript & N/A\tabularnewline
 
Vila and K\"opf, 2017 \cite{VilaK17} & Chromium Linux, Chrome Mac & Shared event loop & JavaScript & $40000$\tabularnewline
 
    \textbf{This work} & \textbf{Chrome (Win, Linux), Firefox (Win, Linux), Safari MacOS, Tor
Linux} & \textbf{Last-level cache} & \textbf{JavaScript} & \textbf{10\textendash500}\tabularnewline
 
\bottomrule
\end{tabular}
\par\end{centering}
}
\end{center}
\end{table*}

Several past works have looked at the possibility of performing website
fingerprinting based on local side-channel information. In all
of these works, which we survey in \cref{t:related-work},
the adversary observes some property of the system while
the victim browser is rendering a webpage. The adversary then applies a machine
learning classifier to the observed side-channel trace to identify
the rendered website.%
  \footnote{
    A different but closely related class of attacks are ``history sniffing''
    attacks, such as \cite{WeinbergCJJ11,LiangYLSH14}, in which the attacker
    wishes to learn which websites the victim has visited in the
  \textbf{past}. 
  }
Some of these works assume that the adversary has malicious control over a
hardware component or
peripheral~\cite{ClarkMRSFX13,LifshitsFHHPTS18,YangGZFB17}.  Others assume that
the adversary can execute arbitrary native code on the target
hardware~\cite{JanaS12a,LeeKKK14,GulmezogluZES17,SpreitzerGKM16}.  
Yet others make the much more modest assumption that the adversary can
induce the victim to render a webpage containing malicious JavaScript 
code~\cite{OrenKSK15,NotSoIncognito,KimL016,VilaK17}.
We mainly investigate the last model.

\citet{KimL016} abuse a data leak in the Chrome implementation
of the Quota Management API, which has been since fixed. Our attack, in
contrast, is based on a fundamental property of the CPU running the browser
application, which is far less trivial to fix. (See \cref{s:countermeasures}.)
Moreover, the mitigations put in place as part of
the response to the Spectre and Meltdown disclosures make the high sampling
rates exploited thus far~\cite{OrenKSK15,VilaK17} unattainable in modern
secure browsers.  Our attack, in contrast, achieves high accuracy at
drastically lower sampling rates and is capable of classifying a significant
number of websites at sampling rates as low as 10\,Hz. To the best of our
knowledge, no cache attack that uses such low clock resolutions has been
demonstrated.

In addition, \citet{OrenKSK15} only recorded a small number  of traces from a few
popular websites, and did not investigate the effectiveness of cache-based
fingerprinting in open-world contexts, or in scenarios where various
anti-fingerprinting measures are in place. 
We address all of these
shortcomings in this work. 
Furthermore, while Oren et al.~\cite{OrenKSK15} do target the Tor Browser, the attack code executes in a
different mainstream browser. Unlike our work, they  do not demonstrate an attack from
JavaScript code running within the Tor Browser.

\citet{NotSoIncognito} is able to classify a
moderate amount of websites using a non-cache-based method with a millisecond clock.
Their attack, however, saturates all of the victim's CPU cores with math-intensive worker
threads, making it highly noticeable and easy to detect by the victim.

\citet{CockGMH14}  implement a covert channel using an L1 cache occupancy channel.
\citet{RistenpartTSS09} show that a cache occupancy channel can detect keystroke timing
and network load in co-located virtual machines on cloud servers.
Both use the technique with high resolution (sub nanosecond) timers.
We are not aware of any prior use of the cache occupancy channel
to overcome low resolution timers.

\section{The Website Fingerprinting Attack Model}\label{s:attack-model}

\begin{figure}[htb]
  \includegraphics[trim={1.5cm 3.5cm 0.5cm 7cm}, clip, page=2,width=\linewidth]{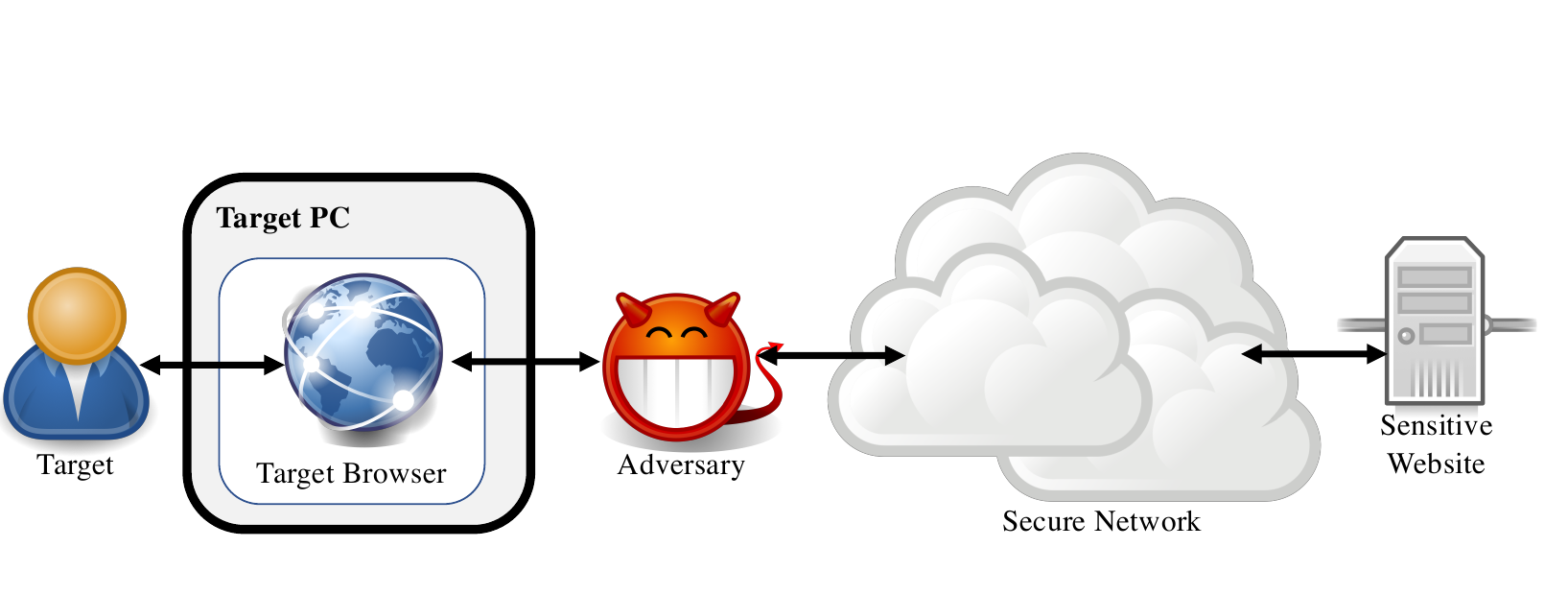}
  \caption{The classical website fingerprinting attack model.
  The (passive) adversary monitors the traffic between the target user and the secure network.
  \label{f:classical-attack-model}}
\end{figure}

The classical attack model used to evaluate website fingerprinting attacks is
presented in \cref{f:classical-attack-model}. In this model, a targeted user uses a
web browser to display a sensitive website. To protect their privacy, the user
does not connect to the website directly, but instead uses a secure network, such as the Tor network,
for the connection.
The attacker is typically modeled as an \emph{on-path adversary}, who is capable of
observing all traffic entering and leaving the Tor network in the direction of the
target user. The adversary cannot understand the contents of the network traffic
since it is encrypted when it enters the Tor network. The adversary is furthermore
unable to directly determine the ultimate destination of the communications after it
exits the Tor network, thanks to Tor's routing protocol.  Finally, due to the
encryption and the validation of the Tor network, the attacker is unable to modify
the traffic without terminating the connection.  An important thread of research on
the security of Tor has investigated the ability of such an adversary to perform
statistical traffic analysis of encrypted traffic, and then to make probabilistic
inferences about users'
communications~\cite{RimmerPJVJ18,PanchenkoLPEZHW16,WangG16,HayesD16,JuarezAADG14,
LuCC10,WangG13,CaiZJJ12,HerrmannWF09,PanchenkoNZE11,Hintz02,YanK18,JansenJGED18,LiGH18}.  
\citet{GongBKS12}
suggest a variation on this scheme, in which the attacker remotely probes routers to
estimate the load of the network traffic they process and performs the statistical
analysis based on this estimated traffic.
\citet{JansenJGED18} suggest another variation in which the attacker monitors the traffic
inside the Tor network, rather then monitoring traffic at the network's edge.

\begin{figure}[htb]
  \includegraphics[trim={3.5cm 3cm 3cm 3cm}, clip, page=4,width=\linewidth]{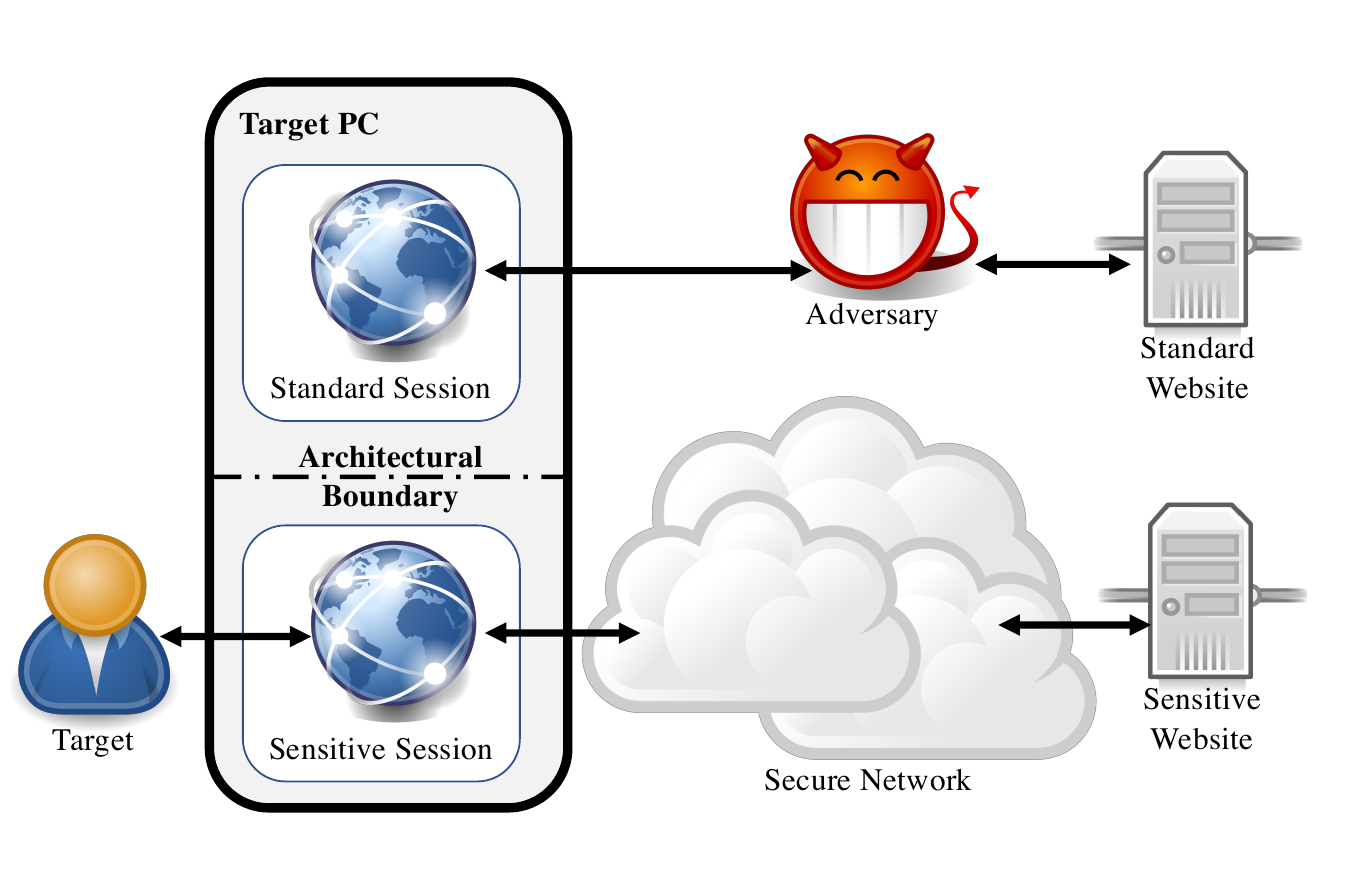}
  \caption{Remote cache-based website fingerprinting attack model. The remote attacker injects malicious JavaScript code into a browser running on the target machine.\label{f:remote-cache-attack-model}}
\end{figure}

In this work we discuss a different attack model, presented in
\cref{f:remote-cache-attack-model}.  In this model, the target user has two concurrent
browsing sessions.
In one session, the user browses to an adversary-controlled site, which
contains some malicious JavaScript code.
In the other session, the user browses to some 
sensitive web site. Due to architectural boundaries, such as sandboxing or process isolation,
the malicious code cannot directly observe the internal state of the sensitive
session.  Hence, the adversary cannot directly determine the ultimate
destination of \emph{any} communication issued from the sensitive session, 
even when the sensitive session
is using a direct unencrypted connection to the remote server.
The malicious code can, however, observe the micro-architectural state
of the processor, and use this information to spy on the sensitive session.

Our attack can therefore be considered in the following scenarios:
\begin{itemize}[itemsep=1ex,topsep=1ex,parsep=0pt,partopsep=0pt]
    \item 
      A \emph{cross-tab scenario}, where a user is made to visit an attacker-controlled website containing malicious JavaScript,
      and this website tries to learn what other sensitive sites the user is visiting at the same time.
      These attacker-controlled and sensitive browsing sessions can be carried out on the same browser,
      on two different browsers belonging to the same user, or even on two 
      browsers residing in two completely isolated virtual machines which share 
      the same underlying hardware~\cite{Qubes}.

      One possible way of causing
      the user to browse to such an attacker-controlled site is through a phishing attack,
      where the attacker sends fraudulent messages, purporting to be from a benign source,
      that induces the victim to click on a link to a malicious web site.  Alternatively,
      the attacker may pay an advertisement service to display a (malicious) advertisement
      when the user visits a third-party website~\cite{GenkinPTY18}. 
      
    \item 
      A \emph{cross-network scenario}, where the attacker is an active on-path adversary  
      capable of injecting JavaScript into any non-encrypted page. The attacker would like to 
      leverage that access to try to learn about the user's sensitive activity, even though 
      the attacker cannot manipulate or access this traffic directly. For example, 
      the user may simultaneously run one browsing session over an unsecured
      connection for mundane tasks, and another browsing session over a second, secured connection
      for sensitive tasks. An attacker capable of modifying traffic on the standard
      link can learn about activity carried out over the secured link, whether this secure connection 
       made through a VPN, through the Tor network, or even through 
      a separate network adapter which the attacker cannot see.

    \end{itemize}
  
The main challenge of the our attack model is
the extremely restricted JavaScript runtime, which requires the attacker code to be
written in a particular way, as we describe further in \cref{s:memorygrammer}.

Regardless of the delivery vector, cache-based fingerprinting has a strong potential
advantage over network-based fingerprinting, since it can indirectly observe both the
computer's network activity and the browser's rendering process. As we demonstrate in
\cref{s:worst-case-molding}, both of these elements contribute to the accuracy of our
classifier.

\section{Data Collection}\label{s:memorygrammer}

\subsection{Creating memorygrams}
The raw data trace for network-based attacks takes the form of a \emph{network
trace}, commonly in the \texttt{pcap} file format, which contains a timestamped
sequence of all traffic observed on a certain network link. The corresponding data
trace in the case of cache attacks is the \emph{memorygram}~\cite{OrenKSK15}---%
a trace of the cache access latency measured at a constant
sampling rate over a given time period. The  memorygrams of 
\citet{OrenKSK15} describe the latency of
multiple individual sets or groups of sets at each point in time, resulting in a
two-dimensional array. 
In contrast, in this work we use a simplified, one-dimensional  memorygram form.
The contents of each entry in our memorygrams is 
a proxy for the occupancy of the cache at the
specific time period.
We collect memorygrams while the browser loads and displays websites,
and use the data as fingerprints for website classification.

\parhead{The Cache Occupancy Channel.}
Unlike prior works~\cite{OrenKSK15,GenkinPTY18}, 
which use the \pp side-channel attack from JavaScript,
we use a cache occupancy channel.
The main difference is that the \pp attack measures 
contentions in specific cache sets,
whereas our attack measures contention over the whole cache.
Specifically, our JavaScript attack allocates an LLC-sized
buffer and measures the time to access the entire buffer.
The  victim's access to memory evicts the contents of our buffer
from the cache, introducing delays for our access.
Thus, the time to access our buffer is roughly proportional to 
the number of cache lines that the victim uses.
Cache occupancy has previously been implemented in native code and used for
covert channels and for measuring co-resident activity~\cite{RistenpartTSS09,CockGMH14}.
Both of these implementations rely on high resolution timers. 
To our knowledge, we are the first to use the cache occupancy channel with a low
resolution timer.

\parhead{Overcoming Hardware Prefetchers.}
Ideally, we would like to collect information across the whole cache.
Intel processors, however, try to optimize memory accesses by prefetching
memory locations that the processor predicts will be accessed in the future.
Because prefetching changes the cache state, we need to fool the prefetchers.
To fool the spatial prefetcher~\cite{IntelOptimizationManual}, we
use the technique of \citet{YaromB14} and do not probe adjacent cache sets.
To fool the streaming prefetcher, which tries to identify sequences of cache
accesses, we use a common approach of masking access
patterns by randomizing the order of the memory accesses we perform~\cite{OsvikST06,LiuYGHL15}.

\parhead{Spatial Information.}
Compared with the \pp attack, the cache occupancy channel does not
provide any spatial information.
That is, the adversary does not learn any information on the addresses
that the victim accesses.
While this is a clear disadvantage of the cache occupancy channel,
our attack does not require spatial information.
The main reason is that modern browsers have complex memory
allocation patterns.
Consequently, the location that data is allocated changes each time
a page is downloaded, and the location carries little information
on the downloaded page.
In practice, not having spatial information is also an advantage.
Without it, there is no need to build eviction sets for cache sets,
a process that can take significant time~\cite{GenkinPTY18}.

\parhead{Website Memorygrams.}
We capture memorygrams when the browser navigates to websites and displays them.  
We use a JavaScript-based memorygrammer to
probe the cache at a fixed rate of one sample every 2\,ms. 
We continue the probe for 30 seconds, resulting in a vector of length 15,000. When a probe 
takes longer than 2\,ms, we miss the slot of the next probe.  We use a
special value to indicate this case.
We use this collection method for all mainstream browsers other
than the Tor Browser,

When the attack code is launched from within the Tor Browser, where the
timer resolution is limited to 100\,ms, we do not measure how long a sweep over the
cache takes, but instead count how many sweeps over the entire cache fit into a
single 100\,ms timeslot. In addition, we do not probe for 30 seconds in this setting,
but rather for 50 seconds, to account for the slower
response time over the Tor network.  Hence, Tor memorygrams contain 500 measurements over 
the entire 50 second measurement time period.

The native code memorygrammer used for the evaluations in \cref{s:robustness} does not suffer from
a reduced timing resolution when measuring the Tor Browser. Therefore, on mainstream browsers 
it runs for 30 seconds and produces 15,000 entries, and on the Tor Browser it runs for 50
seconds and produces 25,000 entries.

\begin{figure}[htb]
    \includegraphics[width=\linewidth]{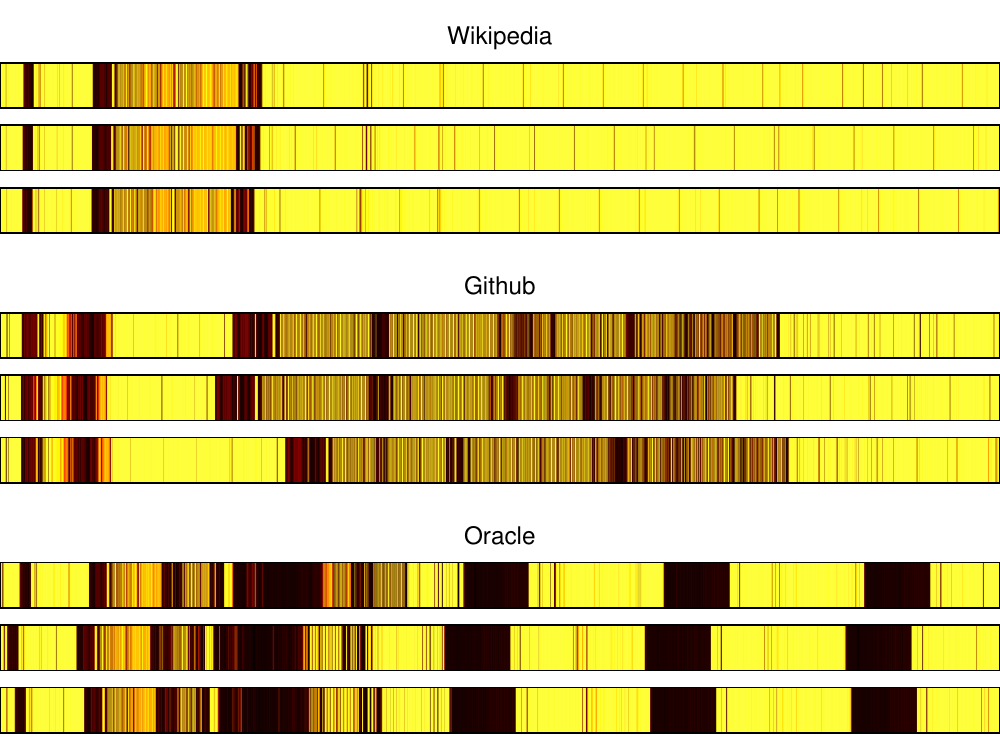}
  \caption{Examples of memorygrams. Time progresses from left to right, shade
  indicates the number of evictions. (Darker shades correspond to more
  eviction.)\label{f:memorygrams}}
\end{figure}

\parhead{Sanity Check.}
Before proceeding, we want to verify that memorygrams can be used for fingerprinting.
Indeed, \cref{f:memorygrams} shows graphical representations of memorygrams of three sites:
Wikipedia (\url{https://www.wikipedia.com}), Github (\url{https://www.github.com}), and
Oracle (\url{https://www.oracle.com}), collected through the native code memorygrammer.
Each memorygram is displayed as a colored strip, 
where time goes from left to right and the shade corresponds to cache activity at each time.
(Lighter shades correspond to fewer evictions.)
We see that the three memorygrams of each site, while not identical, are similar to
each other.
The memorygrams of different websites are, however, very different from each other.
This indicates that memorygrams may be used for identifying websites.

\subsection{Datasets}\label{s:datasets}
\parhead{Closed World Datasets.}
We evaluate our cache-based fingerprinting on six different combinations
of browsers and operating systems, summarized in \cref{tab:JavaScript-code-results}.
Many early works on website fingerprinting operated under a \emph{closed world
assumption}, where the attacker's aim is to distinguish among accesses to a
relatively small list of websites.  Our closed world datasets follow this line
of work.  These datasets consist of 100 traces each for a set of 100 websites,
to a total of 10,000 memorygrams.  We use the same list of 100 websites that
\citet{RimmerPJVJ18} selected from the top Alexa sites.  (See \cref{a:websites}
for a complete list of websites included.) Similar to previous works, no traffic
molding is applied and only one tab is opened at a time. The browser's response cache,
however, is not cleared before accessing each website, an aspect of the experiment we analyze in 
more detail in \cref{s:robustness}.

\parhead{Open World Datasets.}
One common criticism of the closed world assumption is 
that it requires the attacker to 
know the complete set of websites the victim is planning to visit, 
allowing the attacker to prepare and train classifiers for each of these websites. 
This assumption was challenged by many authors, for example \citet{JuarezAADG14}. 
To address this criticism, website fingerprinting methods are 
often evaluated in an open-world setting. 
In this setting, the attacker wishes to monitor access to 
a set of sensitive websites, and is expected to classify them with high accuracy.
Additionally, there is a large set of non-sensitive web pages, 
all of which the attacker is expected to generally label as ``non-sensitive''. 

To evaluate our fingerprinting method in the open-world settings, 
we augment the closed-world datasets
with additional 5,000 traces, each collected for a single unique website, 
again using the list of websites provided by Rimmer et al.~\cite{RimmerPJVJ18}. 
The base rate for this setting is 33.3\%, 
since a trivial classifier can simply decide that all pages are non-sensitive.

\section{Machine Learning}\label{s:ml}
\subsection{Problem Formulation}
Website fingerprinting is generally formulated as a supervised learning problem,
consisting of a template building step and an attack step.  In the template building
step, the adversary visits each target website multiple times and collects a set of
labeled traces (either network traces or memorygrams), each corresponding to a visit
to a certain website. Next, the adversary trains a classifier algorithm on these
labeled traces, using either classical machine learning methods or deep learning
methods. 

In the attack step, the adversary is presented with a set of unlabeled traces, each
one corresponding to a visit to an unknown website. The adversary then applies the
previously trained classifier to each of these traces and outputs a guess for each
trace. The accuracy of the classifier is finally calculated as the percentage of the
correctly assigned labels.

\subsection{Deep Learning Models}\label{s:feature-selection}
Early works on website fingerprinting, starting from \citet{Cheng98trafficanalysis},
used classical machine learning methods such as Naive Bayes, Support Vector Machine
(SVM) and k-Nearest Neighbors (k-NN). As a prerequisite step to running these
classical machine learning methods, the adversary needs to apply an additional
feature extraction step which transforms the raw trace into a more succinct
representation.  Since these features were chosen through human insight into the
nature of network traffic, there was no immediate way of directly applying them to
memorygram analysis.

\citet{abegoto2016} and later \citet{RimmerPJVJ18} suggest
using deep learning for website fingerprinting.
Deep learning performs automatic feature learning from the
raw data, reducing the reliance on human insight at the cost of a larger required
training set. \citet{RimmerPJVJ18} show that, given a large enough training set,
deep-learning website-fingerprinting approaches are as effective as earlier
methods which require
manual feature selection. An advantage of this approach is that it allows us to
compare network-based and cache-based fingerprinting based on the merit of the raw
data, rather than on the specific choice of features.

\parhead{Deep Neural Network Configuration.}
A deep neural network (DNN) is typically
configured as a sequence of non-linear layers which transform the raw data, first
extracting salient features and then selecting the appropriate ones~\cite{9780262035613}. 
Every layer in a
DNN consists of a set of artificial neurons, each connected to a set of
outputs from the previous layers. At the forward propagation stage, 
the activation function is applied to the product of
the each neuron's input and its weight value, and then forwarded to
the next layer. 

For the last layer in the DNNs we evaluate we use a softmax layer,
which outputs a vector containing a-posteriori probabilities for each one of the
classes.

The process of training the neural network uses back-propagation to
update the weights of each
neuron to achieve a minimum loss at the output. 
First, the model calculates the cost between the true
classification of the measurement and the predicted value using a loss function.
Next, the model updates the weights of the each neuron based on the calculated
loss. Every round of forward propagation and back-propagation is called an epoch. A
neural network model runs multiple epochs to learn the weights for accurate
classification.

We evaluate deep learning using two classifier models, Convolutional Neural Networks
(CNN) and Long Short-Term Memory (LSTM) networks~\cite{HochreiterS97}.  A CNN uses
a sequence of feature mapping layers alternating between  
convolutions and max-pooling. Each of the layers
sub-samples the previous layer, iteratively reducing the size of the
input to a more succinct representation, while preserving the information they encode.
Each convolutional layer is a neural network specialised for detecting complex
patterns in its input.  The convolution layer applies several filters to the input
vector, each of which is designed to identify an abstract pattern in a sequence of
input elements it is provided with.  The max-pooling layers reduce the dimensionality
of the data by subsampling the filters, choosing the maximum value from adjacent
groups of neurons applied by the filters.
This alternating sequence of layers extracts complicated features from the
input and produces vectors short enough for the classifiers.  The feature mapping
layers are followed by a \emph{dense} layer, in which every neuron is connected to
every output of the feature extraction phase. The LSTM-based network has an initial feature
selection step similar to the CNN, but then adds an additional layer in which each
neuron has a memory cell, with the output of this neuron determined both by its inputs
and by the value of this memory cell. This allows the classifier to identify patterns
in time-based data.

\parhead{Hyperparameter Selection.}
\emph{Hyperparameters} describe the overall structure of the DNN and of each layer.
The choice of hyperparameters depends on the specific classification problem.
For network-based fingerprinting, we replicated the parameters specified in the dataset
provided by Rimmer et al.~\cite{RimmerPJVJ18}. For cache-based fingerprinting, we manually
evaluated several choices for each hyperparameter. 

To prevent overfitting, we use 10-fold cross validation. We split each dataset
consisting of traces into 10 folds of equal size, and select one fold, consisting
of 10\% of the traces, as a \emph{test set}. The remaining 90\% of the
traces are used for training the classifier, with 81\% serving as the \emph{training set}
and 9\% as the \emph{validation set}. 
The model trains on the training set and the evaluation is done on the test set.
The number of epochs is regulated with an Early-Stop function which stops the
epochs when the accuracy of the validation set no longer increases over successive iterations.
The selected hyperparameters are summarized in~\cref{a:hyperparameters}.

For the CNN classifier we use three
pairs of convolution and max pooling layers.
For the LSTM classifier we use two.
As discussed above, the traces captured by the code running within the
Tor Browser contain only 500 measurements, due to the reduced timer resolution.
For these shorter traces, we modified the architecture of our LSTM-based classifier.
The feature selection of this classifier contains only one convolution layer. We
 therefore used a pool-size of three for the max-pooling layer to limit 
the feature reduction 
before the LSTM layer. In addition, because of the small amount of
features, we could increase the number of LSTM units to 128 and learn more complex patterns
from the features.

\begin{table*}[htb]
  \caption{\label{tab:JavaScript-code-results}Accuracy obtained by in-browser memorygrammer%
      --- Mean (percents) and standard deviation.}
  \begin{centering}
    \begin{tabular}{llllrrrrr}
      \toprule
       Operating &  & LLC & &  \tha{l}{Timer} & \multicolumn{2}{c}{Closed World} & \multicolumn{2}{c}{Open World}\\
       System & CPU &  Size & Browser & \tha{l}{Resolution} &  \multicolumn{1}{c}{CNN} & \multicolumn{1}{c}{LSTM} & \multicolumn{1}{c}{CNN} & \multicolumn{1}{c}{LSTM} \\
      \midrule
      Linux & i5-2500 & 6\,MB & Firefox 59 & 2.0\,ms & 78.5\tpm1.7 & 80.0\tpm0.6 & 86.8\tpm0.9 & 87.4\tpm1.2 \\
      Linux & i5-2500 & 6\,MB & Chrome 64 & 0.1\,ms & 84.9\tpm0.7 & 91.4\tpm1.2 & 84.3\tpm0.7 & 86.4\tpm0.3 \\
      Windows & i5-3470 & 6\,MB & Firefox 59 & 2.0\,ms & 86.8\tpm0.7 & 87.7\tpm0.8 & 84.3\tpm0.6 & 87.7\tpm0.3 \\
      Windows & i5-3470 & 6\,MB & Chrome 64 & 0.1\,ms & 78.2\tpm1.0 & 80.0\tpm1.6 & 86.1\tpm0.8 & 80.6\tpm0.2 \\
      Mac OS & i7-6700 & 8\,MB & Safari 11.1 & 1.0\,ms & 72.5\tpm0.7 & 72.6\tpm1.3 & 80.5\tpm1.0 & 72.9\tpm0.9 \\
      Linux & i5-2500 & 6\,MB & Tor Browser 7.5 & 100.0\,ms & 45.4\tpm2.7 & 46.7\tpm4.1 & 60.5\tpm2.2 & 62.9\tpm3.3 \\
      Linux & i5-2500 & 6\,MB & Tor Browser 7.5 (top 5) & 100.0\,ms & 71.9\tpm2.1 & 70.0\tpm1.7 & 80.4\tpm1.7& 82.7\tpm1.8 \\

      \bottomrule
  \end{tabular}
  \par\end{centering}
  \end{table*}

\section{Results}\label{s:results}

All of the results in this section were obtained by using keras version 2.1.4, with
TensorFlow version 1.7 as the back end, running on two Ubuntu Linux 16.04 servers,
one with two Xeon E5-2660 v4 processors and 128\,GB of RAM, and one with two Xeon
E5-2620 v3 processors and 128\,GB of RAM. Our machine learning instances took
approximately 40 minutes to run in this configuration.

\cref{tab:JavaScript-code-results} presents the fingerprinting accuracy we obtain.
Recall that in this scenario 
the JavaScript interpreter of the targeted browser executes the memorygrammer.
Considering that all modern browsers 
reduced their timer resolution and some added jitter as a countermeasure for 
the Spectre attack~\cite{Pizlo18,Wagner18}, the first question we need to address is
whether it is even possible to implement cache-based fingerprinting attacks in such an environment. 

\begin{figure}[htb]
  \includegraphics[trim={0.5cm 9.5cm 0.5cm 9.5cm},width=\linewidth]{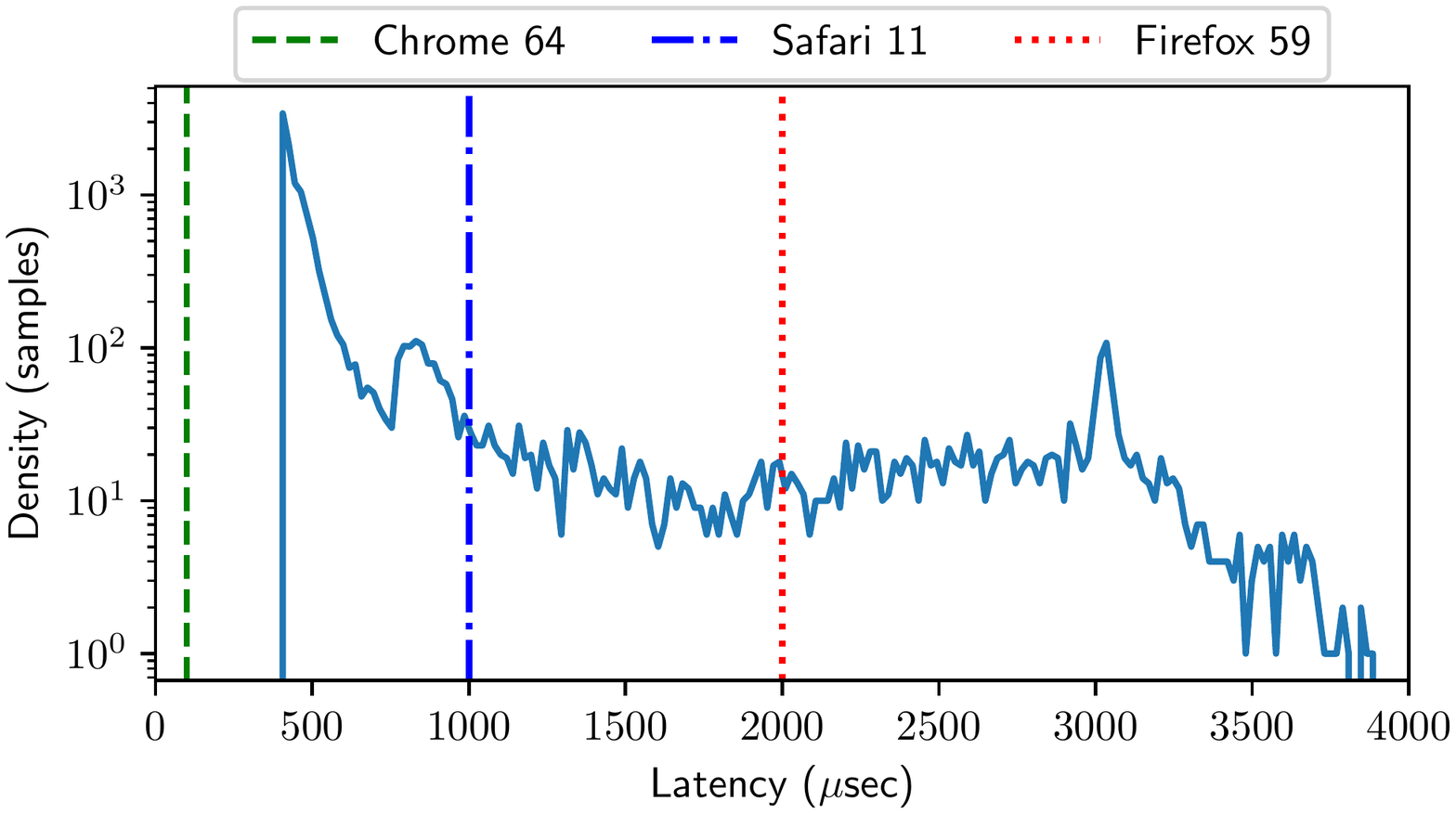}
\caption{Cache probe latencies compared to modern browser timing resolutions.\label{f:full-sum-latencies}}
\end{figure}

To answer this question, we measured the latencies of the cache occupancy
channel using a high-resolution timer while the browser was downloading a web page.
\cref{f:full-sum-latencies} shows the distribution of these latencies.
The figure also uses vertical lines to indicate the timer
resolutions of the various browsers. (See
\cref{tab:JavaScript-code-results}.) 
As we can see, even at the 2\,ms resolution of the Firefox~59 timer, it is
possible to distinguish between 80\% of the probes which take less than 2\,ms
and the remaining 20\%.  This is a welcome side-effect of the use of a large
buffer which is accessed at every probing step.  None of the cache probes we
measured, however, took longer than the 100\,ms clock period of the Tor
Browser.  Hence, when running within the Tor Browser, we count the number
of probes we can perform within each clock tick. (See \cref{s:memorygrammer}.)

The next question is whether the information we collect with this low resolution is
sufficient for fingerprinting.  Indeed, \cref{tab:JavaScript-code-results} shows that
in all of the environments we test our classifier is significantly better than a
random guess.
Remarkably, as our results show, even the highly restricted Tor
Browser can be used for mounting cache attacks, albeit with a
significantly lower accuracy than that of general-purpose browsers.

\subsection{Closed World Results}\label{s:closed-world}
We first look at the typical closed-world scenario investigated by past works.
In mainstream browsers, our JavaScript attack code is consistently able to
provide classification accuracies of 70--90\%, well over the base rate of 1\%.
The Tor Browser attack, however, achieves a lower accuracy of 47\%. If we,
however, look not only at the top result output by the classifier, but also
check whether the correct website is one of the top 5 detected websites, the
accuracy of the Tor Browser attack climbs to 72\%, with a base rate of 5\%.
This method of looking at the few most probable outputs of a classifier was
previously used in similar classification
problems~\cite{Caliskan-IslamHLNVYG15,NarayananPGBSSS12}.
With some a-priori information an attacker can deduce which
of the top 5 pages the victim has accessed.

We can compare the accuracy of our cache-based fingerprinting to the one obtained by 
state-of-the-art network-based methods, as reported by~\citet{RimmerPJVJ18}. 
We see that while there are differences between the classification accuracy achieved in each
case, the overall accuracy is comparable, assuming both attacks capture the same amount of
traces per website. 
As in the network-based setting, we believe that capturing more than 100 traces per website
is likely to increase the accuracy and the stability of our classifier.

\subsection{Open World Results}\label{s:open-world}

We next turn to the more challenging open-world scenario, in which the 100
sensitive webpages must be distinguished from an additional set of 5,000
non-sensitive pages. As seen in \cref{tab:JavaScript-code-results} the
JavaScript-based website fingerprinting code performs well under this scenario
as well, again achieving classification accuracy of 70--90\%. We note
that in most cases the results are slightly better than the closed-world
results.  The reason is the larger size of the  ``non-sensitive'' class.  As
discussed earlier, this also significantly increases the base rate for
open-world scenarios to 33.3\%.

As in the case of the closed-world setting, we can evaluate the accuracy of the Tor Browser
under a top-5 assumption, i.e.\ when checking for the correct website in the
top five outputs of the classifier.
Under this relaxation the Tor Browser attack achieves a high 
accuracy rate of 83\%, with a base rate of 37.3\%.

The classification to sensitive vs.\ non-sensitive site is a binary classification problem, 
We can, therefore,  apply standard analysis techniques to this aspect of the results.
We achieved a near perfect classification in all of the open world settings we evaluated, 
achieving an area under curve (AUC) of more than 99\% in all cases.

\begin{table*}[htb]
  \begin{center}
  \caption{\label{tab:robustness-results}Accuracy obtained in robustness tests %
      --- Mean (percents) and Standard deviation.}
  \begin{tabular}{lrrrrrrrr}
  \toprule
        & \multicolumn{2}{c}{Firefox Network} & \multicolumn{2}{c}{\textbf{Firefox Cache}} 
        & \multicolumn{2}{c}{Tor Network} & \multicolumn{2}{c}{\textbf{Tor Cache}} \\
    Test& \multicolumn{1}{c}{CNN} & \multicolumn{1}{c}{LSTM} 
	& \multicolumn{1}{c}{\textbf{CNN}} & \multicolumn{1}{c}{\textbf{LSTM}}
        & \multicolumn{1}{c}{CNN} & \multicolumn{1}{c}{LSTM} 
	& \multicolumn{1}{c}{\textbf{CNN}} & \multicolumn{1}{c}{\textbf{LSTM}}\\
  \midrule
    Baseline & 86.4\tpm1.0 & 93.2\tpm0.5 & \textbf{94.9\btpm0.5} & \textbf{94.8\btpm0.5}
             & 77.6\tpm1.6 & 90.9\tpm0.7 & \textbf{72.7\btpm0.7} & \textbf{80.4\btpm0.5}\\

    Response cache enabled & 56.1\tpm1.5 & 70.6\tpm1.5 & \textbf{92.2\btpm0.8} & \textbf{92.2\btpm0.5}
                           & 55.5\tpm1.7 & 65.9\tpm1.0 & \textbf{86.1\btpm0.5} & \textbf{86.3\btpm0.6}\\
    Render only & \tha{c}{--} & \tha{c}{--} & \tha{c}{--} & \tha{c}{--} 
                & 1.0\tpm0.0 & 1.0\tpm0.0& \textbf{63.3\btpm1.1} & \textbf{63.9\btpm1.5}\\
    Network only & \tha{c}{--} & \tha{c}{--} & \tha{c}{--} & \tha{c}{--} 
		 & 77.6\tpm1.6 & 90.9\tpm0.7 & \textbf{19.9\btpm1.8} & \textbf{51.9\btpm2.7}\\
    Temporal drift & \tha{c}{--} & \tha{c}{--} & \tha{c}{--} & \tha{c}{--} 
		&  64.5\tpm2.2 & 81.0\tpm0.6 & \textbf{68.3\btpm0.5} & \textbf{75.6\btpm0.7} \\
    
    \bottomrule
  \end{tabular}
  \end{center}
\end{table*}

\section{Robustness Tests\label{s:robustness}}
Having demonstrated the effectiveness of our website fingerprinting technique,
we now turn our attention to its robustness and test its resilience to issues
known to affect network-based fingerprinting. 

\subsection{Evaluation Setup}
\begin{figure}[htb]
    \includegraphics[trim={0.5cm 4cm 1.3cm 4cm},clip, page=3,width=\linewidth]{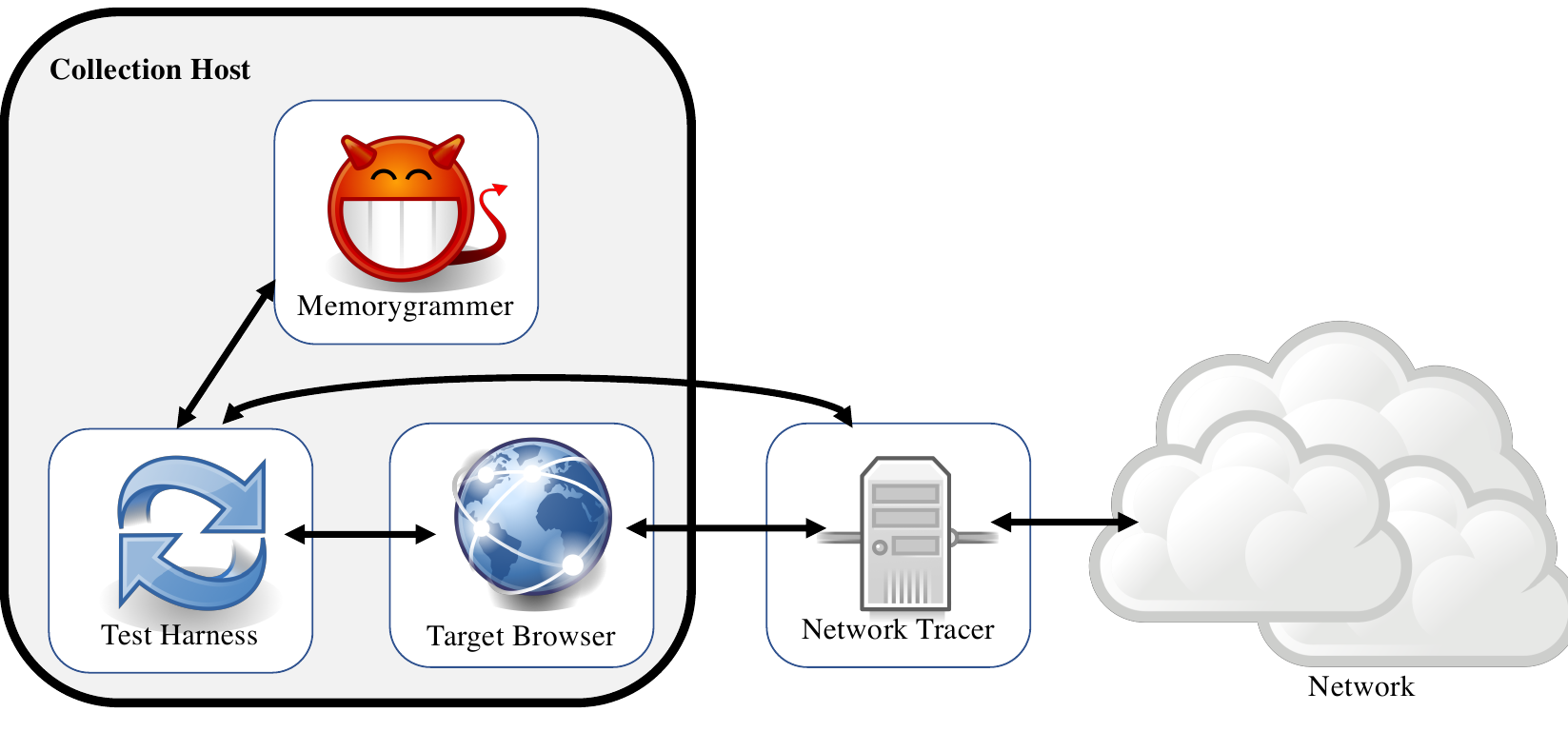}
  \caption{Data Collection Setup for the Robustness Tests.\label{f:robustness-setup}}
\end{figure}

To compare the results of network fingerprinting with cache-based 
fingerprinting, we need to modify our data collection setup.  The setup,
illustrated in \cref{f:robustness-setup}, consists of two data collection
hosts.  The \emph{memorygram collection host}, which simulates the victim's
machine, runs both the target browser and the memorygrammer software.  The
\emph{network tracer} sits on-path between the memorygram collection hosts and
the Internet, and collects a record of the network traffic. A test harness
written in Perl and Python invokes the memorygrammer, the network tracer and
the target browser at the same time, then saves a correlated data record
consisting of the memorygram, the network trace in \texttt{pcap} format, and a
screenshot of the target web page for monitoring purposes.
For data collection, we use 
HP Elite 8300 desktop
computers featuring Intel Core i5-2500 CPUs at 3.30\,GHz, with a 6\,MB
last-level cache, running CentOS 7.2.1511 and either Firefox 59 or Tor Browser 7.5.

For the robustness tests we use a native-code memorygrammer,
which is based on the \pp implementation of Mastik, a
side-channel toolkit released under the GNU Public License~\cite{Yarom16}.
We apply two modifications to the Mastik code.
First, we change the \pp code to measure cache occupancy rather than 
activity in specific cache sets.
Secondly, we use the processor's performance counters~\cite{IntelDevelopersManual}
to count the number of cache evictions rather than use the high resolution timer
to identify evictions.
The use of
performance counters for attack purposes has already been proposed and investigated
in the past~\cite{UhsadelGV08,BhattacharyaM15,LeeSGKKP17,BrasserMDKCS17}.

\subsection{Baseline Scenario}
Our baseline scenario replicates the results of our closed world JavaScript
memorygrammer, as well as some of the results of \citet{RimmerPJVJ18}.
As we can see in \cref{tab:robustness-results}, 
the native-code memorygrammer gives a slightly better accuracy than the JavaScript
memorygrammer on Firefox.
When attacking the Tor Browser, the native code memorygrammer achieves much
better results than the in-browser JavaScript code.
We believe that the cause of the improvement is the higher probing accuracy
afforded by the native-code memorygrammer.
In both browsers, the results of the native-code memorygrammer are similar to
those achievable with network-based fingerprinting.

\subsection{Enabling the Response Cache}\label{s:keep-cache}

Network-based fingerprinting methods, by definition, 
must rely on network traffic to perform classification. 
Typically, due to caching, many 
web pages are loaded with partial or no network traffic. 
As specified in RFC 7234~\cite{RFC7234}, the performance of web browsers is 
typically improved by the use of response caches.  
When a web browser client requests a remote resource from a web server, 
the server can specify that a particular response is cacheable, 
and the web browser can then store this response locally, either on disk or in memory. 
When the page is next requested, 
the web browser can ask the server to send the response  
only if it has been modified since the last time it was accessed by the client. 
In the case of a response cache hit, 
the server only returns a short header instead of the complete remote resource, 
resulting in a very short network traffic sequence. 
In some cases, the client can even reuse the cached 
response without querying the server for a remote copy, 
resulting in no network traffic at all. 
\citet{HerrmannWF09} demonstrate a significant decrease
in  the accuracy of 
web fingerprinting when the browser uses the response cache. 
Indeed, deleting or disabling the browser cache prior
to fingerprinting attacks is a common practice~\cite{WangG13,PanchenkoNZE11}.

We enable caching of page contents by the browser, and measure the effect
on fingerprinting accuracy.
In the Firefox browser we simply refrain from clearing the response
cache between sessions.
For privacy reasons, the response cache in the Tor Browser
does not persist across session restarts. 
Hence, when collecting data on the Tor Browser
we ``prime'' the cache before every recording by opening the web page in another tab, 
allowing it to load for 15 seconds, then closing the tab.

When we keep the browser's response cache, 
the advantage of cache-based website fingerprinting starts to emerge.
As  \cref{tab:robustness-results} shows, the accuracy of the 
standard network-based methods degrades when the response caching is enabled. 
We can see a degradation in accuracy of over 20\% in the fingerprinting accuracy.
 
In contrast, the cache-based methods are largely unaffected by the 
reduction in network traffic, 
achieving high accuracy rates.
This result supports the conclusion that the cache-based detection methods 
are not simply detecting the CPU activity related to the handling of network traffic, 
making them essentially a special case of network-based classifiers, 
but are rather detecting rendering activities of the browser process.

\subsection{Net-only and Render-only Results}\label{s:worst-case-molding}
\citet{OrenKSK15} show that cache activity is correlated with network activity,
raising the possibility that cache-based fingerprinting basically identifies
the level of network activity.  To rule out this possibility and show that
website rendering also contributes to fingerprinting, 
we separate rendering (or more precisely, data processing) activity 
from handling of network data.

\parhead{Render-Only Fingerprinting.}
To capture the data processing activity,
we neutralize the network activity by guaranteeing constant traffic levels.
More specifically, we apply molding to the network traffic,
ensuring that data flow between the collection host and the network at a fixed
bandwidth of 10\,KB every 250\,ms.
To achieve that, we queue data transmitted at a higher rate, or send
dummy packets when the transmitted data does not fill the desired bandwidth.
These dummy packets are silently dropped by the receiver.
The approach is, basically, BuFLO~\cite{DyerCRS12}, with $\tau=\infty$,
i.e., when the data stream continues indefinitely.
This approach 
has a high bandwidth overhead compared to WTF-PAD and WT,
however, it is designed to ensure that the network traffic is constant
irrespective of the contents of the website.
As expected, the raw network
captures in this scenario all have the exact same size,
which happens to be twice as
large as the largest network capture recorded without traffic molding.

Because all the traces are identical, the network-based classifier assigns the
same class to all of the traces, and its accuracy is the same as a random
guess.  The results of cache-based fingerprinting show a drop in accuracy
compared with unmolded traffic. However, the accuracy is still significantly
better than a random guess.  This experiment demonstrates the resilience of
cache-based website fingerprinting to mitigation techniques aimed at
network-based fingerprinting, and suggests that this privacy threat should be
countered using a different class of mitigation techniques, as we explore
further in \cref{s:countermeasures}.

\parhead{Network-Only Fingerprinting.}
In a complementing experiment, we aim to capture only the network
traffic.
To collect this dataset, we 
first capture actual traffic data from a real
browsing session. 
We then use a mock setup, that does not involve a browser at all.
Instead, 
we use two \texttt{tcpreplay}~\cite{tcpreplay} instances, one at the collection host,
and the other at a server, to emulate the network traffic,
by replaying the data from the \texttt{pcap} file.

The results for this experiment show that the cache-based
classifier is capable of classifying many pages even when no rendering
activity is taking place. However, the accuracy is significantly lower than
in the case that rendering activity does take place.
In particular, our CNN classifier only detects the correct website
in about 20\% of the cases, significantly lower than the 73\% we get
for the matching closed-world scenario. 
(But still much better than the 1\% expected for a random guess.)
The accuracy of the network-based
classifier  is the same as for the baseline, simply because
the network traffic is replicated.

Combining these two experiments we therefore conclude that cache-based fingerprinting
identifies features both in the network traffic patterns and in the actual
\emph{contents} of the displayed web pages.

\subsection{Dealing with Temporal Drift}\label{s:drift}
The accuracy of network-based website
fingerprinting decays over time, when the contents of the website changes~\cite{RimmerPJVJ18}.
Many
websites use content management systems (CMS), in which the page layout is based on a fixed
template design, and only the resources loaded into this template vary over
time. Since, as we have shown, the cache-based fingerprints capture rendering
activities as well as network activities, it would seem that the rendering-related
traces recorded by the cache-based method would have a longer lifetime, and be more
resistant to drift, than the network-related traces captured by the traditional
method. 

To test this hypothesis, we repeat the data collection of the baseline experiment
after a delay of 36 days (start to start). We then measure the ability of both
cache-based and network-based classifiers to accurately classify the new traces,
after being trained on the old traces. 
In this setting, we see a drop of 5--10\% in the  accuracy of both classifiers.
We believe that further experiments are required for accurately assessing
how cache-based and network-based fingerprinting handle temporal drifts.

\section{Detecting Unknown Hardware Configurations}\label{s:probe-hardware}
In contrast to network-based fingerprinting, which is largely target agnostic,
cache-based fingerprinting needs to be tailored to the precise hardware configuration
of the victim machine, specifically the set count and associativity of its last-level
cache. 
Using a too large or a too small buffer reduces the 
effectiveness of the  technique, and eventually the
accuracy of the classifier. There are, however, not that many popular configurations. 
For example, four cache configurations (4096 or 8192 sets, 12 or 16 ways) cover most 
of the Intel Core processor models.

If the target hardware configuration is known
beforehand (assuming, for example, that a particular user is singled out for attack)
the attacker can customize the parameters of the JavaScript attack code to match the
target PC's parameters. It would be interesting, however, to see how well an attacker
can remotely determine an unknown target's cache configuration using JavaScript. To
investigate this, we created a JavaScript program that allocates a 20MB array in
memory and iterates over it in several patterns which should fit in well into
different configurations of cache set-counts and associativities. We then recorded the
minimum, maximum and mean access time per element, plus the standard deviation, for
each of these configurations. We collected 1,350 such measurements from multiple
systems with cache sizes of 3\,MB, 4\,MB, 6\,MB, and 8\,MB. 
We then used
MATLAB's classification learner tool to apply a variety of machine learning
classifiers to the measured data. Using both KNN and SVM classifiers, we were able to
correctly classify the configuration of the target's last-level cache with over
99.8\% classification accuracy under 5-fold cross validation. Interestingly, even a
simple tree-based classifier which compared the minimum iteration time of three
different configurations to a predefined threshold was 99.6\% accurate. We ported
this simple tree-based classifier to JavaScript, creating an LLC cache size detector
which we tested and found capable of accurately detecting the cache sizes of 15
different machines with diverse browser, hardware and operating system
configurations, taking less than 300\,ms to run in all cases. 
Thus generic attacks that adapt to the specific hardware configuration seem feasible.

  \begin{figure*}[htb]
    \includegraphics[width=\linewidth]{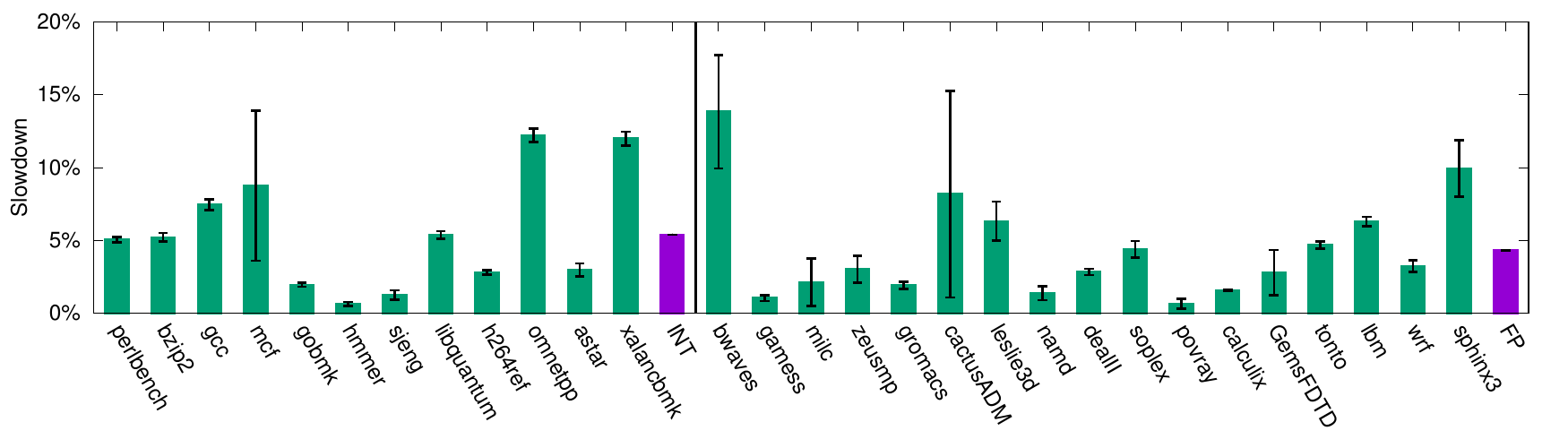}
    \caption{Performance slowdown of our countermeasure on the SPEC benchmark. Error bars indicate one standard deviation. INT and FP show the geometric mean of the SPEC integer and floating point benchmarks, respectively.\label{f:specbar}}
\end{figure*}

\section{Countermeasures}\label{s:countermeasures}
We now discuss potential countermeasures to our fingerprinting attack.
We first describe a cache masking technique we experimented with. We then  follow with
a review of other cache attack countermeasures suggested in the literature.

\subsection{Cache Activity Masking}

One well-studied mitigation method from the domain of network-based cache fingerprinting
involves creating spurious network activity to mask the actual website traffic~\cite{DyerCRS12}.
It is possible to adapt such a masking technique to our domain and
mask the actual website rendering activity by creating spurious activity in the cache.  
Our initial experiments show that this is a promising
mitigation, but further research is needed to assess its effectiveness and its
effect on performance and on power consumption.

\parhead{Masking implementation.}
Our countermeasure repeatedly evicts the entire last-level.
More specifically, we allocate a cache-sized
buffer and access every cache line in the buffer in a loop.  Such masking could be applied in the browser,
in the operating system, as a browser plugin, and
even incorporated into a security-conscious website in the form of JavaScript
delivered to the client. For our initial proof of concept implementation we chose to implement the 
countermeasure as a standalone native code application, based on a modification of the Mastik
side-channel toolkit~\cite{Yarom16}. This setting allows us to investigate the effectiveness
of our countermeasure while leaving deployment complexities for future work.

\parhead{Evaluation.}
We evaluated this countermeasure on a desktop computer featuring an Intel Core
i5-2500, running Centos Linux version 7.6.1810.  We enabled the countermeasure,
then collected website traces both for Firefox (Linux) and for the Tor Browser,
using the same mix of traces described in \cref{s:datasets}---10,000 traces
for the closed-world scenario, consisting of 100 traces for each of the Alexa
top 100 websites, and 5,000 additional traces for the open-world scenario, each
collected for a single unique website. We split the data set into training,
testing and validation sets and applied 10-fold cross validation, as described
in more detail in \cref{s:feature-selection}.

Our experiments show that the countermeasure completely thwarts the attack when
training is done on an unprotected system---the accuracy of our classifier was
at or below the base rate of 1\% for the closed-world scenario and 33\% for the
open-world scenario.  We also evaluated a scenario in which the adversary is
allowed to train on traces with the countermeasure applied. In this more
challenging scenario, the countermeasure completely thwarts the attack when the
attack code is running from the Tor Browser. On Firefox, however, we only
noticed a moderate reduction in the effectiveness of the attack.  In the closed
world scenario, the attack achieves 73\% success and in the open world the
success rate is 77\%. (Down from 79\% and 86\%, respectively.)

\parhead{Performance Impact.}
To understand the effect that our countermeasure has system performance, we used the industry-standard SPEC CPU
benchmark~\cite{specCPU}, the de-facto standard benchmark for measuring the performance of the CPU and the memory subsystems.
\cref{f:specbar} shows the results of the SPEC CPU 2006 benchmarks with 
our countermeasure, relative to no countermeasure.
The countermeasure causes a slowdown
of around 5\% (geometric mean across the benchmarks) with a worst case slowdown of 14\%
for the \textsf{bwaves} benchmark.
 These
results are from the average of ten executions of the benchmarks for each case.
With Tor network performance being as it is, we believe that the performance
hit on CPU benchmarks is acceptable for this scenario.

\subsection{Other Countermeasures}
Most of the past research into cache attacks has been done in the context
of side-channel cryptanalysis. 
Due to the different scenario, many of the countermeasures typically suggested
for cache-based attack are no longer effective.
Techniques such as constant-time programming~\cite{BernsteinLS12} are only applicable
to regular code, typically found in implementations of cryptographic
primitives.
It is hard to see how such techniques can be applied to web browsers.
Similarly, as this work demonstrates, timer-based defenses that reduce the timer
frequency or add jitter are not effective.

Cache randomization techniques~\cite{Qureshi18,LiuL14,WangL07} dissociate victim and adversary cache sets,
and prevent the adversary from monitoring victim access to specific addresses.
However, our attack measures the overall cache activity rather than looking 
at specific victim accesses.
As such, such techniques are unlikely to be effective against
our attack.

Cache partitioning, either using dedicated hardware~\cite{WangL07,DomnitserJLAP12} or via page coloring~\cite{LiedtkeHH97},
is a promising approach for mitigating cache attacks.
In a nutshell, the approach partitions the cache between security domains,
preventing cross-domain contention.
Web pages are often rendered within the same browser process.
A page-coloring countermeasure will, therefore, need to adapt to the browser scenario.
Alternatively, the current shift to strict site isolation~\cite{SiteIsolation} as part of the mitigations
for Spectre~\cite{kocher2018spectre}, may assist in applying page coloring to protect against our attack.
A further limitation of page coloring is that caches support only a handful of
colors.
Hence, colors need to be shared, particularly when a large number of tabs are open.
To provide protection, page coloring will have to be augmented
with a solution that prevents concurrent use of the same color by
multiple sites.

\textsc{CacheBar}~\cite{ZhouRZ16} limits the contention caused by each process as a protection
for the \pp attack. 
Like cache partitioning, this approach works at a process resolution and
may require adaptions to work in the web browser scenario. 
Furthermore, unlike past cryptographic attacks that aim to identify specific
memory accesses, our technique measures the overall memory use of the victim.
Consequently, unless \textsc{CacheBar} is configured to partition the cache,
some cross-process contention will remain, allowing our attack to work.

\section{Limitations and Future Work}
While the work demonstrates the feasibility of cache-based website fingerprinting and
provides an analysis of the attack, it does leave some areas for further
study.  Being the first analysis of its kind, 
the scope of the work does not match
the scope of similar works on network-based website fingerprinting.  In particular,
our datasets are significantly smaller than those of \citet{RimmerPJVJ18}, for
example.  Providing larger datasets would allow better analysis of the effectiveness
of the technique and would be a beneficial service for the research community as a
whole.

In this work we collected the memorygrams on the same hardware configuration used by
the victim PC.  While we show that we can adapt the data collection to the specific
victim hardware (\cref{s:probe-hardware}), 
at this stage it is not clear how much a classifier trained on data
collected with one hardware configuration would be effective for classifying
memorygrams collected on a different configuration.

In the network-based
website fingerprinting scenario, little to no traffic travels through the network unless the user is actively
fetching a webpage. In the cache-based scenario, however, the cache is always active to a degree, even before the 
browser starts to receive and render the webpage. Recognizing the start of a trace may therefore
be more difficult in the cache-based setting than in the network-based setting, especially in the case of a
real attack.
Our framework implicitly synchronizes the trace with the start of the download.
Due to varying network conditions, we see differences of up to six seconds
between trace start and render start. As such, we believe that our technique
can identify web sites even without the synchronization.  Further
experimentation is required, however, to verify this fact.
We also note that if the machine is otherwise idle, cache activity can serve as a (slightly noisy) indicator
of the start of the trace.

The work further shares many of the limitations of network-based
fingerprinting~\cite{JuarezAADG14}.  In particular, websites tend to change over time
or based on the identity of the user or the specifications of the computer used for
displaying them.  Furthermore, our work, like most previous works, assumes that only
one website is displayed at each time.  
Both \citet{RimmerPJVJ18} and our work
briefly discuss
temporal aspects of website fingerprinting, and we also looked a bit into
the issue (\cref{s:drift}).  However, further work is required to
assess the impact of this and other variables on the efficacy of cache-based
fingerprinting.

\section{Conclusions}
In this work we investigate the use of cache side channels for website fingerprinting.
We implement two memorygrammers, which capture the cache activity of the browser,
and show how to use deep learning to identify websites based on the cache activity 
that displaying them induces.

We show that cache-based website fingerprinting achieves results comparable with 
the state-of-the-art network-based fingerprinting.
We further show that cache-based fingerprinting outperforms network-based fingerprinting
under a common operating scenario, where the browser maintains cached objects.
Finally, we demonstrate that cache-based fingerprinting is resilient to both traffic
molding and to reduced timer resolution.
The former being the standard defense for network-based website fingerprinting
and the latter the currently implemented countermeasure for 
mobile-code-based microarchitectural attacks.
To the best of our knowledge, this is the first cache-based side channel attack that works
with the 100\,ms clock rate of the Tor Browser.

\ifAnon
\else
\section*{Acknowledgements}
We would like to thank Vera Rimmer for her helpful comments and insights.
We would also like to thank Roger Dingledine and our shepherd Rob Jansen for reviewing and commenting on the final version of this paper.

This research was supported by 
the ARC Centre of Excellence for Mathematical \& Statistical Frontiers,
Intel Corporation,
Israel Science Foundation grants 702/16 and 703/16,
NSF CNS-1409415, and NSF CNS-1704105.
\fi
\ifCCS
  \bibliographystyle{ACM-Reference-Format}
\fi
\ifUSENIX
  \bibliographystyle{plainnat}

\fi
{\footnotesize 
\setlength{\bibsep}{3pt plus 2pt minus 2pt}
\bibliography{cacheprint.bib}
}

\newpage

\begin{appendices}

  \crefalias{section}{appsec}
  \section{Selected Hyperparameters}\label{a:hyperparameters}
Tables~\ref{t:hyperparameters-cnn}, \ref{t:hyperparameters-lstm}, and \ref{t:hyperparameters-tor}
summarize the hyperparameters for the classifiers used in this work.

\begin{table}[htb]
  {\footnotesize 
\centering
  \caption{Hyperparameters for the CNN classifier\label{t:hyperparameters-cnn}}
\begin{tabular}{lll}
\toprule
  \textbf{Hyperparameter} & \textbf{Value} & \textbf{Space} \\
\midrule
Optimizer          & Adam   & Adamax, Adam, SGD, RMSprop \\
Learning rate      & 0.001  & 0.001--0.002 \\
Batch size         & 100    & 40--100      \\
Training epoch     & 20--30 & Early stop by accuracy       \\
Convolution layers & 3      & 3--4         \\
Input units (FF)   & 15000  & 15000--25000 \\
Input units (Tor)  & 25000  & 15000--25000 \\
CNN activation     & relu   & relu, tanh   \\
Kernels            & 256    & 2--512       \\
Kernel size        & 16,8,4 & 2--31        \\
Pool size          & 4      & 2--8         \\
\bottomrule
\end{tabular}
  }
\end{table}

\begin{table}[htb]
  {\footnotesize
\centering
  \caption{Hyperparameters for the LSTM classifier\label{t:hyperparameters-lstm}}
  \begin{tabular}{lll}
\toprule
  \textbf{Hyperparameter} & \textbf{Value} & \textbf{Space}\\
  \midrule
Optimizer                    & Adam  & Adamax, Adam, SGD, RMSprop \\
Learning rate                & 0.001 & 0.001--0.002 \\
Batch size                   & 100   & 40--100      \\
Training epoch               & 20--30     & Early stop by accuracy           \\
Convolution layers           & 2     & 1--3         \\
Input units (FF)             & 15000 & 15000--25000 \\
Input units (Tor)            & 25000 & 15000--25000 \\
CNN activation               & relu  & relu, tanh   \\
LSTM activation              & tanh  & relu,tanh    \\
Kernels                      & 256   & 2--512       \\
Kernel size                  & 16,8  & 2--32        \\
Pool size                    & 4     & 2--8         \\
Dropout                      & 0.2   & 0.1--0.2     \\
LSTM units                   & 32    & 8,32        \\
\bottomrule
\end{tabular}
  }
\end{table}

\newpage

\begin{table}[t]
  {\footnotesize
\centering
  \caption{Hyperparameters for the LSTM classifier for the Tor attack\label{t:hyperparameters-tor}}
  \begin{tabular}{lll}
\toprule
  \textbf{Hyperparameter} & \textbf{Value} & \textbf{Space}\\
  \midrule
Optimizer                    & Adam  & Adamax, Adam, SGD, RMSprop \\
Learning rate                & 0.001 & 0.001--0.002 \\
Batch size                   & 100   & 40--100      \\
Training epoch               & 20--30     & Early stop by accuracy           \\
Convolution layers           & 1     & 1--3         \\
Input units                  & 500   & 500          \\
CNN activation               & relu  & relu, tanh   \\
LSTM activation              & tanh  & relu,tanh    \\
Kernels                      & 256   & 2--512       \\
Kernel size                  & 32  & 2--32        \\
Pool size                    & 3     & 2--8         \\
Dropout                      & 0.4   & 0.1--0.4     \\
LSTM units                   & 128   & 8,32,128     \\
\bottomrule
\end{tabular}
  }
\end{table}

  \section{Websites Included in Closed-World Datasets}\label{a:websites}
  {

\begin{tabular}{ll}
9gag.com & abs-cbn.com \\
adf.ly & adobe.com \\
aliexpress.com & allegro.pl \\
amazon.com & amazonaws.com \\
aol.com & apple.com \\
archive.org & askcom.me \\
battle.net & blastingnews.com \\
booking.com & breitbart.com \\
bukalapak.com & businessinsider.com \\
conservativetribune.com & dailymail.co.uk \\
dailymotion.com & detik.com \\
deviantart.com & dictionary.com \\
digikala.com & doubleclick.net \\
doublepimp.com & ebay.com \\
espncricinfo.com & exoclick.com \\
extratorrent.cc & facebook.com \\
\end{tabular}
\newpage
\begin{tabular}{ll}
feedly.com & gamepedia.com \\
github.com & go.com \\
godaddy.com & goodreads.com \\
google.com & hclips.com \\
hola.com & hotmovs.com \\
imdb.com & instructure.com \\
intuit.com & kompas.com \\
leboncoin.fr & liputan6.com \\
livejasmin.com & livejournal.com \\
ltn.com.tw & microsoftonline.com \\
mozilla.org & msn.com \\
naver.com & netflix.com \\
nicovideo.jp & nih.gov \\
ntd.tv & office.com \\
onedio.com & openload.co \\
oracle.com & ouo.io \\
outbrain.com & pinterest.com \\
popads.net & quora.com \\
researchgate.net & roblox.com \\
rt.com & rutracker.org \\
scribd.com & skype.com \\
soundcloud.com & sourceforge.net \\
spotify.com & spotscenered.info \\
stackexchange.com & stackoverflow.com \\
steamcommunity.com & steampowered.com \\
t.co & theguardian.com \\
thesaurus.com & tistory.com \\
tokopedia.com & torrentz2.eu \\
tribunnews.com & tumblr.com \\
twitter.com & weather.com \\
wikia.com & wikipedia.org \\
wittyfeed.com & xhamster.com \\
xvideos.com & yandex.ru \\
yelp.com & zippyshare.com \\
\end{tabular}
}

\end{appendices}

\end{document}